\documentclass[aps,prd,preprint,preprintnumbers,nofootinbib,superscriptaddress]{revtex4-2}
\usepackage{amssymb}
\usepackage{amsmath}
\usepackage{amsfonts}
\usepackage{graphicx}
\usepackage{booktabs}
\usepackage[format=hang,justification=RaggedRight]{caption}
\usepackage{subcaption}

\newcommand{\Eqref}[1]{Eq.~\eqref{#1}}

\newcommand{\ctc}{\cos{\vartheta_\text{coll}}}
\newcommand{\stch}{\sin\left(\frac{\vartheta_{\rm coll}}{2}\right)}
\newcommand{\ctch}{\cos\left(\frac{\vartheta_{\rm coll}}{2}\right)}
\newcommand{\stchq}{\sin^2\left(\frac{\vartheta_{\rm coll}}{2}\right)}
\newcommand{\ctchq}{\cos^2\left(\frac{\vartheta_{\rm coll}}{2}\right)}
\newcommand{\st}{\sin{\vartheta}}
\newcommand{\ct}{\cos{\vartheta}}
\newcommand{\sph}{\sin{\varphi}}
\newcommand{\cph}{\cos{\varphi}}

\allowdisplaybreaks

\begin{document}

\setlength{\unitlength}{1mm}

\title{Two-beam laser photon merging} 

\author{Chantal Sundqvist}\email{chantal.sundqvist@uni-jena.de}
\affiliation{Helmholtz-Institut Jena, Fr\"obelstieg 3, 07743 Jena, Germany}
\affiliation{GSI Helmholtzzentrum f\"ur Schwerionenforschung, Planckstra\ss e 1, 64291 Darmstadt, Germany}
\affiliation{Theoretisch-Physikalisches Institut, Abbe Center of Photonics, \\ Friedrich-Schiller-Universit\"at Jena, Max-Wien-Platz 1, 07743 Jena, Germany}
\author{Felix Karbstein}\email{felix.karbstein@uni-jena.de}
\affiliation{Helmholtz-Institut Jena, Fr\"obelstieg 3, 07743 Jena, Germany}
\affiliation{GSI Helmholtzzentrum f\"ur Schwerionenforschung, Planckstra\ss e 1, 64291 Darmstadt, Germany}
\affiliation{Theoretisch-Physikalisches Institut, Abbe Center of Photonics, \\ Friedrich-Schiller-Universit\"at Jena, Max-Wien-Platz 1, 07743 Jena, Germany}

\date{\today}

\begin{abstract}
Quasi-elastic scattering processes have long been thought of providing the most promising signal for a first experimental detection of quantum vacuum nonlinearity. A prominent example of such a process is vacuum birefringence. However, these signals are typically strongly background dominated. This problem can be circumvented by inelastic scattering processes.
In this study, we investigate the inelastic process of laser photon merging in the collision of just two laser pulses under a finite angle, which provides signal photons of a distinct frequency outside the frequency spectrum of the background.
As a key result, for the example of two laser beams of the same oscillation frequency we demonstrate that by using high-intensity optical lasers and choosing an optimal collision angle, photon merging should become accessible in experiment with state-of-the-art technology. In this case three frequency $\omega$ laser photons are merged to a single $3\omega$ photon.
\end{abstract}

\maketitle

\newpage

\section{Introduction}
The nature of the quantum vacuum is governed by quantum fluctuations. In the case of quantum electrodynamics (QED), these allow electromagnetic fields to interact nonlinearly by the coupling to virtual electron-positron pairs \cite{Euler:1935zz,Heisenberg:1935qt,Weisskopf} (for recent reviews, see Refs.~\cite{DiPiazza:2011tq,Dunne:2012vv,Battesti:2012hf,King:2015tba,Inada:2017lop,Karbstein:2019oej,Fedotov2022}). However, these effective couplings are parametically suppressed by powers of $|\vec{E}|/E_{\rm cr}$ and $|\vec{B}|/B_{\rm cr}$ with the critical electric (magnetic) field strength $E_{\rm cr}=m_e^2c^3/(e\hbar) \simeq 1.3 \times 10^{18}\,\text{V}/\text{m}$ ($B_{\rm cr}=E_{\rm cr}/c\simeq 4\times 10^9\,\text{T}$). The strongest macroscopic electromagnetic fields presently available in the laboratory are generated by high power lasers reaching $|\vec{E}|\approx \mathcal{O}(10^{14})\,\text{V}/\text{m}$ and $|\vec{B}|\approx \mathcal{O}(10^6)\,\text{T}$ in $\mu$m-sized focal volumes, such that generically $|\vec{E}|\ll E_{\rm cr}$, $|\vec{B}|\ll B_{\rm cr}$. These circumstances have so far prevented the direct observation of quantum vacuum signatures under controlled laboratory conditions.
With ongoing advances in laser technology and the building of new dedicated high-intensity laser facilities, a particularly promising route to an experimental verification of QED vacuum nonlinearity is provided by all-optical pump-probe type setups.
The attainable photonic signatures in this type of experiment can be divided in two main classes, namely {\it quasi-elastic} and manifestly {\it inelastic} processes.

{\it Quasi-elastic} processes depend only on the oscillation frequency of one of the driving beams; in the monochromatic plane-wave limit they become strictly elastic.
This results in signal photons with kinematic properties very similar to the probe photons.
A prominent example of such a process is vacuum birefringence \cite{Toll:1952,Baier,BialynickaBirula:1970vy,Brezin:1971nd, Heinzl2006, DiPiazza2006}.
Generically, {\it quasi-elastic} processes provide satisfactory large signal photon numbers which however contend with the large background of the driving lasers.\\
For laser fields which can be modeled as paraxial beams {\it inelastic} processes depend on the frequencies of both lasers.
Typically, inelastic signatures are suppressed relatively to elastic ones.
On the upside, the emission direction as well as the energy of the signal photons arising from inelastic scattering processes often differ significantly from those constituting the driving laser beams.
Examples of inelastic signatures of quantum vacuum nonlinearity are photon splitting \cite{BialynickaBirula:1970vy,Adler:1970gg,Adler:1971wn,Papanyan:1971cv,Stoneham:1979,Baier:1986cv,Adler:1996cja,DiPiazza:2007yx} and photon merging \cite{Bialynicka-Birula1981, Rozanov:1993, Kaplan2000, Valluri2003, DiPiazza2005, Marklund2006, Fedotov2007, Narozhny2007, Yakovlev:1966,DiPiazza:2007cu,Gies:2016czm, Huang2019, Sasarov2021}.

So far, the great potential of inelastic quantum vacuum signatures for all-optical experiments has mainly been exemplified in scenarios involving multiple ($>2$) or specially tailored laser beams, cf., e.g. \cite{Rozanov:1993,Mckenna:1963,Varfolomeev:1966,Moulin:2002ya,Lundstrom:2005za,Gies:2016czm,Boehl2015,Gies:2017ezf,King:2018wtn,Karbstein:2019dxo}.
The availability of just two fundamental-frequency high-intensity laser beams is typically considered as insufficient to achieve sizable inelastic signals in experiment.
In the present work, we provide a thorough analysis of the effect of laser photon merging in the collision of two identical laser pulses at zero impact parameter.
To this end, we analyze the emission characteristics of the merged signal photons in a scenario envisioning the collision of two pulsed, paraxial Gaussian laser beams under a finite angle.

Our paper is organized as follows: in Sec.~\ref{sec:scenario} we briefly recall the theoretical foundations and detail the analytical modeling of our specific setup. This provides us with an analytic expression for the differential number of signal photons which we will use in Sec.~\ref{sec:signal_charcterization} to deduce the emission characteristics of the merging signal. In Sec.~\ref{sec:results} we provide explicit results for the angular distribution and the total number of merged signal photons attainable in a polarization insensitive measurement. Finally, we end with Conclusions and an Outlook in Sec.~\ref{sec:concls}.

\section{Theoretical foundations}
\label{sec:scenario}
Our analysis is based on the vacuum emission picture \cite{Galtsov:1971xm,Karbstein:2014fva}, which allows to recast all-optical signatures of quantum vacuum nonlinearity in prescribed macroscopic electromagnetic fields as signal photon emission processes.
The leading processes are zero-to-single signal photon transitions. The central object thereby is the zero-to-single signal photon transition amplitude $\mathcal{S}_{(p)}(\vec{k})$ to a state with one signal photon of wave-vector $\vec{k}$, energy $k^0=|\vec{k}|$ and polarization $p$.
It is related to the differential number of signal photons to be measured far outside the interaction region of the driving laser fields as
	\begin{equation}
		{\rm d}^3N_{(p)}=\frac{{\rm d}^3k}{(2\pi)^3}\bigl|{\cal S}_{(p)}(\vec{k})\bigr|^2\,.
		\label{eq:d3N}
	\end{equation}
Regarding the study of quantum vacuum nonlinearities, the currently attainable laser fields can be considered as locally constant, weak fields, i.e. fields that vary on spatial scales much larger than the Compton wavelength of the electron $\lambdabar_{\rm C}=\hbar/(m_ec) \simeq 3.86\times 10^{-13}\,{\rm m}$ and fulfill $\{|\vec{E}|,c|\vec{B}|\}\ll E_{\rm cr}$. A thorough derivation of the signal photon transition amplitude at one-loop order recapitulating particularly the approximations made for locally constant, weak fields can be found in Ref.~\cite{Gies:2017ygp}. All considerations presented are based on the leading correction term to classical Maxwell theory $\mathcal{L}_{\rm int}\sim 4\mathcal {F}^2 + 7\mathcal{G}^2$ with the field invariants $\mathcal{F} = (\vec{B}^2 - \vec{E}^2)/2$ and $\mathcal{G} = -\vec{E}\vec{B}$.

\begin{figure}
			\centering
			\includegraphics[width=0.85\textwidth]{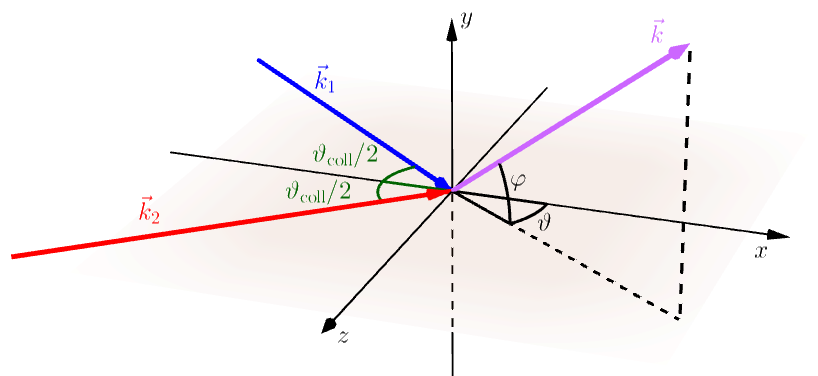}
			\caption{Sketch of the collision geometry. The wave vectors of the two colliding laser pulses are $\vec{k}_1$, $\vec{k}_2$. The collision takes place in the $xz$-plane under a collision angle $\vartheta_{\rm coll}$. Its bisector is identified with the x-axis. The wave vector of the signal photons is $\vec{k}$ and parameterized by the angles $\varphi$, $\vartheta$.}
			\label{fig:3omega_CollisionGeometrySketch}
		\end{figure}

\begin{figure}
		\centering
		\includegraphics[width=0.4\textwidth]{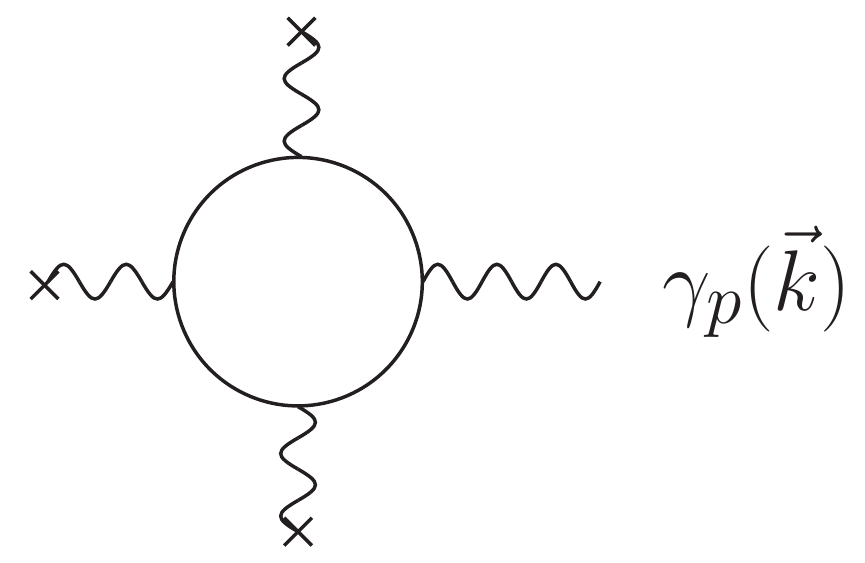}
		\caption{Feynman diagram of the leading vacuum fluctuation induced corrections to classical Maxwell theory in the limit of weak fields, interpreted as a vacuum emission process. The crosses ``$\times$'' mark couplings to the laser fields. The depicted process results in signal photons $\gamma$ with wave vector $\vec{k}$ and polarization $p$.}
		\label{fig:Feynman_merg_simple}
	\end{figure}

Here, we study the process of laser photon merging in the collision of two identical, paraxial laser beams with linear polarization.
Without loss of generality, we choose these to collide in the $xz$-plane with the $x$-axis as the bisecting line of the collision angle $\vartheta_{\rm coll}$. See Fig.~\ref{fig:3omega_CollisionGeometrySketch} for a sketch of the collision geometry.
The unit wave vectors of the beams $b \in \{1, 2 \}$ are $\hat{\vec{k}}_1 = \left(\cos\left( \frac{\vartheta_{\rm coll}}{2} \right),0,\sin\left( \frac{\vartheta_{\rm coll}}{2} \right) \right)$ and $\hat{\vec{k}}_2 = \left( \cos\left( \frac{\vartheta_{\rm coll}}{2} \right),0,-\sin\left( \frac{\vartheta_{\rm coll}}{2} \right) \right)$. As polarization vectors we use $\hat{\vec{e}}_{\beta_1}=\left( -\sin\left( \frac{\vartheta_{\rm coll}}{2} \right)\cos\beta_1, \sin\beta_1,\cos\left( \frac{\vartheta_{\rm coll}}{2} \right)\cos\beta_1 \right)$ and $\hat{\vec{e}}_{\beta_2}=\left( \sin\left( \frac{\vartheta_{\rm coll}}{2} \right)\cos\beta_2, \sin\beta_2,\cos\left( \frac{\vartheta_{\rm coll}}{2} \right)\cos\beta_2 \right)$. A particular choice of the angle parameter $\beta_b$ fixes the polarization of beam $b\in\{1,2\}$. The associated electric and magnetic fields are given by $\vec{E}_b = \mathcal{E}_b \hat{\vec{E}}_b$ and $\vec{B}_b = \mathcal{E}_b \hat{\vec{B}}_b$ with the amplitude profile $\mathcal{E}_b$. They fulfill $\hat{\vec{E}}_b \perp \hat{\vec{B}}_b \perp \hat{\vec{k}}_b$.
We parameterize the wave vectors of the signal photons by $\hat{\vec{k}}=\left(\cos{\varphi}\cos{\vartheta},\sin{\varphi},\cos\varphi\sin{\vartheta}\right)$  with $-\frac{\pi}{2}\leq \vartheta \leq \frac{\pi}{2}$ and $-\pi \leq \varphi \leq \pi$. For $\varphi=0$, $\vartheta=0$ this matches the bisector of the collision angle $\vartheta_{\rm coll}$ between the incident beams. As polarization vector of the signal photons we use $\hat{\vec{e}}_\beta=\left(-\cos\beta\sin\vartheta-\sin\beta\sin\varphi\cos\vartheta, \sin\beta\cos\varphi,-\sin\beta\sin\varphi\sin\vartheta+\cos\beta\cos\vartheta \right)$.
\\Using these notations and labeling the signal polarizations by the angle parameter $\beta$, ${\cal S}_{(p)}(\vec{k})\to{\cal S}_\beta(\vec{k})$, we obtain
	\begin{equation}
		\begin{aligned}
			{\cal S}_{\beta}(\vec{k})=&\ {\rm i} \frac{8}{45} \frac{\alpha^2}{m_e^4} \sqrt{\frac{k^0}{2}} \stchq \sum_{m=1}^2 {\cal I}_{m,3-m}(\vec{k}) \\
			& \times \left\{ \left[\cph-\cos\left(\vartheta - (-1)^m \frac{\vartheta_{\rm coll}}{2} \right) \right] f\left(\beta_1+\beta_2,\beta+\beta_{3-m}\right) \right.\\
			& \hspace{3cm}\left. - \sph \sin\left( \vartheta - (-1)^m \frac{\vartheta_{\rm coll}}{2} \right) f\left( \beta_1+\beta_2,\beta+\beta_{3-m}+\frac{\pi}{2} \right) \right\} \, ,
		\end{aligned}
	\label{eq:transitionamplitude}
	\end{equation}
where we have made use of the shorthand notations $f(\mu,\nu)=4\cos\mu\cos\nu+7\sin\mu\sin\nu $ and
\begin{equation}
	{\cal I}_{mn}(\vec{k})=\int{\rm d}^4 x\,{\rm e}^{{\rm i} k^0(\hat{\vec 
			k}\cdot\vec{x}-t)}\,{\cal E}_1^m(x){\cal E}_2^n(x)\,. \label{eq:Imn}
\end{equation}
The leading contribution to the signal photon emission from the laser-driven QED vacuum amounts to a vacuum-fluctuation-mediated four-field interaction; see Fig.~\ref{fig:Feynman_merg_simple}.
One of these four fields is the signal photon field, induced by the effective interaction of three laser fields.
A single paraxial field does not induce signal photons.
Consequently, the powers of the field profiles ${\cal E}_b$ in \Eqref{eq:Imn} in a two-beam collision are limited to $(m,n)\in\{(1,2),(2,1)\}$. \\
Aiming at an analytic evaluation of \Eqref{eq:Imn}, we model the fields of the driving lasers as paraxial Gaussian pulses in the infinite Rayleigh range approximation, \cite{Gies:2017ygp,King:2018wtn, Karbstein:2019oej},
	\begin{equation}
		{\cal E}_b(x)={\frak E}_b\,{\rm e}^{-\left(\frac{\vec{x}\cdot\hat{\vec{\kappa}}_{b}-t}{\tau_b/2}\right)^2} {\rm e}^{-\frac{\vec{x}^2-\left(\vec{x}\cdot\hat{\vec{\kappa}}_{b}\right)^2}{w_{0,b}^2}} \cos\left\{\omega_b\left(\vec{x}\cdot\hat{\vec{\kappa}}_{b}-t\right)\right\} \, . \label{eq:E}
	\end{equation}
This approximation is well justified as long as the spatial extents of the interaction region of the laser beams are governed by a length scale much smaller than the Rayleigh range $z_{R,b}=\frac{w_{0,b}^2 \omega_b}{2}$.
For collision angles fulfilling $\left\{\frac{w_{0,1}}{z_{R,2}} , \frac{w_{0,2}}{z_{R,1}}\right\} \lesssim \sin\vartheta_{\rm coll}$ this is ensured automatically; cf. the detailed discussion in \cite{Karbstein:2021}. This corresponds to a constraint to $18.6^\circ \ll \vartheta_{\rm coll} \ll 161.4^\circ$ for diffraction limited beams with $w_0 \approx \lambda$, i.e. $w_0 \omega =2 \pi$. For less tightly focused laser beams, this approximation can be applied to a wider range of collision angles.\\
The profiles~\eqref{eq:E} are chosen such that both beams $b\in\{1,2\}$ reach the peak field amplitude $\frak{E}_b$ in their common beam focus at $\vec{x}=0$ at exactly the same time.
The peak field amplitude is related to the laser pulse energy $W_b$, pulse duration $\tau_b$ and waist size $w_{0,b}$ as \cite{Karbstein:2017jgh}
	\begin{equation}
		{\frak E}_{b}\simeq 2\Bigl(\frac{8}{\pi}\Bigr)^{\frac{1}{4}}\sqrt{\frac{W_b}{\tau_b w_{0,b}^2\pi}}\,.
		\label{eq:PeakFieldAmplitude} 
	\end{equation}
Note, that $\tau_b$ and $w_{0,b}$ are measured at $1/{\rm e}^2$ of the peak intensity. The conversion of these parameters into quantities at half maximum (HM) is carried out according to
\begin{equation}
	\tau_b^{\rm HM}=\tau_b\sqrt{\frac{\ln 2}{ 2}} \qquad \text{and} \qquad w_{0,b}^{\rm HM}=w_{0,b} \sqrt{\frac{\ln 2}{2}} \, .
\end{equation}

Performing the Fourier integration in Eq.~\eqref{eq:Imn} for the field profiles \eqref{eq:E} with equal laser parameters, i.e. $\tau_1=\tau_2\equiv\tau$, $w_{0,1}=w_{0,2}\equiv w_0$, $\omega_1 = \omega_2 \equiv \omega$ and $W_1=W_2\equiv W$, we obtain 
	\begin{equation}
		\begin{aligned}
		{\cal I}_{mn}(\vec{k})=& \left(\frac{\pi}{2}\right)^2 \left(\frac{{\frak E}}{2}\right)^{m+n} \frac{w_0^3 \tau^2}{(m+n)\sqrt{m n H}\stch} \sum_{l=0}^m \sum_{j=0}^n \binom{m}{l} \binom{n}{j} {\rm e}^{-\frac{w_0^2(k^0)^2}{4(m+n)}\frac{h_+ h_-}{\stchq}} \\
		&\times   {\rm e}^{-\frac{\tau^2}{16 (m+n)}\left[k^0+(m-2l+n-2j)\omega\right]^2} {\rm e}^{-\frac{k^0 \omega \tau^2 w_0^2}{4 H (m+n)}(h_+ + h_- -2\stchq)(2j-n+2l-m)} \\
		&\times {\rm e}^{-\frac{w_0^2 \tau^2}{16 m n (m+n)H\stchq}\left[4\omega(mj-nl)\stchq+k^0(h_- m -h_+ n)\right]^2} \label{eq:Imn_explicit}
		\end{aligned}
	\end{equation}
with
	\begin{equation}
		h_\pm = \cph\cos\left( \vartheta \pm \frac{\vartheta_{\rm coll}}{2} \right)- 1
		\label{eq:hpm}
	\end{equation}
and
	\begin{equation}
		H=\tau^2 \ctchq +4 w_0^2 \stchq \, .
		\label{eq:H}
	\end{equation}\\
We can straightforwardly infer from the exponential factors in Eq.~\eqref{eq:Imn_explicit} that for sufficiently large pulse durations $\{\omega_1\tau_1,\omega_2\tau_2\}\gg1$ as considered throughout this work, the energy of the signal photons is essentially determined by the oscillation frequencies/ photon energies of the driving laser beams. 
In particular, the integers $m$ and $n$ count the number of couplings of the fields of beam 1 and beam 2 to the electron positron loop, respectively, and the integers $l$ and $j$ specify whether the laser fields absorb or release energy. Here the following cases are possible: For $l=0$ and $j=0$ energy is absorbed by the respective laser field. Conversely, for $l=m$ and $j=n$ energy is released by the laser field. Finally, for either $(m,l)=(2,1)$ or $(n,j)=(2,1)$ one of the beams does effectively not influence the energy of the signal photons because the same energy is absorbed at one coupling and released at the other coupling of the same field. Hence, these cases describe quasi-elastic scattering processes where the signal photon energy depends only on the photon energy of one of the beams.\\
The absolute square of the transition amplitude, Eq.~\eqref{eq:transitionamplitude}, required in the calculation of the differential number of signal photons, Eq.~\eqref{eq:d3N}, can be expressed as a sum of terms proportional to $\mathcal{I}^*_{m_1n_1} \mathcal{I}_{m_2n_2}$ with $(m_i,n_i)\in\{(1,2),(2,1)\}$, i.e. ${\rm d}^3 N \sim |\mathcal{I}_{12} + \mathcal{I}_{21}|^2 = |\mathcal{I}_{12}|^2 + |\mathcal{I}_{21}|^2 + \mathcal{I}_{12}\mathcal{I}_{21}^* + \mathcal{I}_{12}^*\mathcal{I}_{21}$. Based on the fact that for slowly varying pulse envelopes as considered here the energy dependence in Eq.~\eqref{eq:Imn_explicit} is essentially described by $\mathcal{I}_{mn}\sim \exp{\left\{-\frac{\tau^2}{16(m+n)}\left[k^0 +(m-2l+n-2j)\omega\right]^2\right\}}$, this leads to the condition
	\begin{align}
		\left[k^0 +(m_1-2l_1+n_1-2j_1)\omega\right]^2 + \left[k^0 +(m_2-2l_2+n_2-2j_2)\omega\right]^2 = 0
	\end{align}
to yield a sizeable signal. This results in the following energy for the signal photons
	\begin{align}
		k^0 \simeq& \omega\left[ l_1+j_1+l_2+j_2-3 \pm {\rm i} \left(l_2+j_2-l_1-j_1\right) \right]\, . \label{eq:k0}
	\end{align}
Of course, the signal photon energy $k^0$ has to take on a real, positive value. We therefore conclude that the dominant signals are encoded either in contributions with $(l_i,j_i) \in \{(0,2),(2,0),(1,1)\}$ which lead to signal photons of energy $k^0 \approx \omega$, or in $(l_i,j_i)=(m_i,n_i) \in \{(1,2),(2,1)\}$ leading to $k^0 \approx 3\omega$.

\section{$3\omega$-signal}
\label{sec:signal_charcterization}
In the remainder of this work, we will exclusively focus on the contributions to Eq.~\eqref{eq:Imn_explicit} which induce a signal at a photon energy $k^0 \simeq 3\omega$ and refer to this as the $3\omega$-signal. This signal is particularly interesting because the signal photon energy lies outside of the frequency spectrum of the driving lasers. On a microscopic level, two frequency $\omega$ photons of one laser beam merge together with a frequency $\omega$ photon of the other \textit{assisting} laser beam to form a single outgoing signal photon. For completeness, we note that this process persists in the zero frequency limit for the assisting field, where two photons of the same beam merge to yield a $2\omega$ signal. \\

As the $3\omega$-signal photons clearly differ from the background of the frequency $\omega$ laser photons in energy, they should be discernible in experiment without the immediate need to consider additional quantum vacuum induced modifications of the signal such as polarization. Therefore, we sum over the two transverse polarization states characterized by polarization vectors with $\beta$ and $\beta+\frac{\pi}{2}$ resulting in the polarization insensitive differential number of signal photons with $k^0 \approx 3\omega$,
	\begin{equation}
		\begin{aligned}
			\frac{{\rm d}^3 N^{3\omega}}{{\rm d}k^0 {\rm d}\varphi {\rm d}\!\sin\vartheta} =& \frac{\pi}{8} \left(\frac{7}{45}\right)^2 \frac{\alpha^4}{m_e^8} \stchq \frac{(k^0)^3 w_0^6 \tau^4}{9 H} \left(\frac{{\frak E}}{2}\right)^6 \left(16 + 33 \sin^2\left(\beta_1+\beta_2\right)\right)\\
			&\hspace{-0.3cm}\times \sum_{p=1}^2 \sum_{q=1}^2 \Bigg\lbrace  \bigg[  \prod_{m=p,q}\left(\cph - \cos{\left(\vartheta-(-1)^m \frac{\vartheta_{\rm coll}}{2}\right)} \right)  \\
			&\qquad \quad+\sin^2\varphi \prod_{m=p,q} \left(\sin{\left(\vartheta-(-1)^m \frac{\vartheta_{\rm coll}}{2}\right)}\right)\bigg] \cos\left(\beta_{3-p}-\beta_{3-q}\right)  \\
			&\quad+ 2 \sph \stch h (q-p)\sin\left(\beta_{3-p}-\beta_{3-q}\right)  \Bigg\rbrace \\
			&\hspace{-0.3cm}\times  {\rm e}^{-\frac{\tau^2}{24} (k^0-3\omega)^2} {\rm e}^{-\frac{w_0^2 (k^0)^2 \sin^2\varphi}{12}} {\rm e}^{-\frac{2(k^0)^2 w_0^4 h^2}{3H}} \\
			&\hspace{-0.3cm}\times  {\rm e}^{-\frac{(k^0)^2 w_0^2 \tau^2}{96 H \stchq}\left[9(h_-^2 + h_+^2) +16 \stchq (h_+ + h_- +\stchq)+3\Theta_{pq} (h_-^2 - h_+^2) \right]} \, ,
		\end{aligned}
		\label{eq:d3N_3omg}
	\end{equation}
where
	\begin{equation}
		h=\cph \ct - \ctch
		\label{eq:h}
	\end{equation}
and
	\begin{equation}
		\Theta_{pq} = \left\{\begin{array}{lr}
			-1, & \text{for } p=q=1\\
			0, & \text{for } p\neq q \\
			1, & \text{for } p=q=2
		\end{array}\right. \, .
	\end{equation}
Here, we used the notation $\prod_{m=p,q}c_m=c_p c_q $.\\
The terms with $p=q$ correspond to the two \textit{direct} contributions ($\sim \left|\mathcal{I}_{12}\right|^2$ and $\sim \left|\mathcal{I}_{21}\right|^2$, respectively) and the two terms with $p\neq q$ are the interference terms or \textit{indirect} contributions ($\sim \mathcal{I}_{12}\mathcal{I}_{21}^*$ and $\sim \mathcal{I}_{12}^*\mathcal{I}_{21}$, respectively). The direct contributions to the differential number of $3\omega$-signal photons are invariant under the transformations $\vartheta_{\rm coll}\rightarrow -\vartheta_{\rm coll}$ and $\vartheta\rightarrow -\vartheta$ for the specific setup considered.
Furthermore, the differential number of signal photons associated with the direct terms, i.e. $p=q$, depends on the polarization of the incident beams only via the overall prefactor $16+33\sin^2(\beta_1+\beta_2)$. In order to maximize the signal, the two polarizations should thus be related via $\beta_1+\beta_2=\frac{\pi}{2}$. In addition to the same overall factor, for the interference terms, i.e. $p\neq q$, there appear additional factors depending on the polarization of the incident beams. Interestingly, these render the optimal choice for $\beta_1$, $\beta_2$ dependent on both the collision angle and the signal photon emission direction encoded in $\varphi$, $\vartheta$ and $\vartheta_{\rm coll}$. However, because of the symmetry of the considered collision geometry, we can safely assume, that the signal's maximum lies in the collision plane where $\varphi=0$. Upon insertion of $\varphi=0$ into Eq.~\eqref{eq:d3N_3omg}, the directional and polarization dependences factorize. We conclude that $\cos(\beta_p-\beta_q)$ should be maximized to maximize the signal photon numbers. This leaves us with the two conditions $\beta_1+\beta_2=\frac{\pi}{2}$ and $\beta_p-\beta_q=0$ to be simultaneously fulfilled to yield the largest signal. From these it is easy to infer that the interference term contributes most to the signal for $\beta_1=\beta_2=\frac{\pi}{4}$. Being exclusively interested in the maximum signal we will adopt this choice for the polarizations of the incident beams in the remainder of this article.\\

\subsection{Emission characteristics}
In a next step we aim at deriving relatively simple analytical scalings. To perform the integration over $k^0$ in Eq.\eqref{eq:d3N_3omg} we use the fact that in the parameter regime of interest to us the signal is strongly peaked at $k^0=3\omega$; see also Refs.~\cite{Karbstein:2018omb,Karbstein:2019oej}: first, we identify all factors of $k^0$ in the prefactor to the exponential functions in \Eqref{eq:d3N_3omg} with $k^0=3\omega$. Second, we formally extend the integration limits to $\pm \infty$, such that $\int{\rm d}k^0\to \int_{-\infty}^\infty{\rm d} k^0$. The resulting integral of Gaussian type can be readily integrated analytically and be expressed in terms of elementary functions.
Therewith, we obtain the following (approximate) expression for the emission-angle resolved differential signal photon number
\begin{equation}
	\begin{aligned}
		&\frac{{\rm d}^2 N^{3\omega}}{{\rm d}\varphi {\rm d}\!\sin\vartheta} \approx \frac{(3\pi)^\frac{3}{2}}{\sqrt{2}}\left(\frac{7}{45}\right)^2 \frac{\alpha^4}{m_e^8} \sin^3\left(\frac{\vartheta_{\rm coll}}{2}\right) \left(\frac{\mathfrak{E}}{2}\right)^6 \frac{\omega^3 w_0^6 \tau^4}{H} \sum_{p,q=1,2}\\
		&\times \left[\prod_{m=p,q}\left(\cph - \cos{\left(\vartheta-(-1)^m \frac{\vartheta_{\rm coll}}{2}\right)} \right) +\sin^2\varphi \prod_{m=p,q} \left(\sin{\left(\vartheta-(-1)^m \frac{\vartheta_{\rm coll}}{2}\right)}\right) \right] \\
		&\times \Biggl\{4\tau^2 \stchq + 8w_0^2 \left[\sin^2\varphi \stchq + 2h^2\right] \\
		&\hspace{3cm}+ \frac{w_0^2 \tau^2}{H}\left[9(h_-^2 + h_+^2)-4(h_- + h_+)^2 +3\Theta_{pq}(h_-^2 - h_+^2)\right]\Biggr\}^{-\frac{1}{2}} \\
		&\times {\rm e}^{-\frac{3\omega^2\tau^2}{8} \frac{8w_0^2 \left[\sin^2\varphi \stchq + 2h^2\right] + \frac{w_0^2 \tau^2}{H}\left[9(h_-^2 + h_+^2)-4(h_- + h_+)^2 +3\Theta_{pq}(h_-^2 - h_+^2)\right]}{4\tau^2 \stchq + 8w_0^2 \left[\sin^2\varphi \stchq + 2h^2\right] + \frac{w_0^2 \tau^2}{H}\left[9(h_-^2 + h_+^2)-4(h_- + h_+)^2 +3\Theta_{pq}(h_-^2 - h_+^2)\right]}}
	\end{aligned}
	\label{eq:d2N_3omega}
\end{equation}
with $h_\pm$, $H$ and $h$ as introduced in Eqs.~\eqref{eq:hpm}, \eqref{eq:H} and \eqref{eq:h}.
A sizable signal can only be generated if the exponential suppression is minimized. For the $3\omega$-signal as presented in Eq.~\eqref{eq:d2N_3omega} the exponential suppression is minimized for co-propagating beams, i.e. $\vartheta_{\rm coll}=0$. This is also in line with plane wave considerations, \cite{Gies2021}. However, for paraxial beams, no signal photons can be generated for this collision angle because $\mathcal{F} = \mathcal{G} = 0$, which manifests itself in the prefactor of Eq.~\eqref{eq:d2N_3omega}. Thus, we have to bear in mind that considering only the exponent of the expression for the differential number of signal photons \eqref{eq:d2N_3omega} is insufficient to infer the properties of the signal photons' emission characteristics. However, since the exponential suppression outweighs the influence of the prefactor, we can still say, without having a quantitative prediction at this stage, that the optimal collision angle $\vartheta_{\rm coll}^{\rm max}$, i.e. the collision angle yielding the maximum signal, must be rather small (especially compared to the best choice of $\vartheta_{\rm coll}^{\rm max}=\pi$ for the frequency $\omega$ signal).\\
From Eq.~\eqref{eq:d2N_3omega} we can furthermore corroborate what we expected from the system's symmetry: firstly that the signal is maximal in the collision plane, i.e. for $\varphi^{\rm max}=0$, and secondly that the signal photons originating from the interference terms are primarily emitted at $\vartheta=0$, which is the bisector of the collision angle $\vartheta_{\rm coll}$.\\
On the other hand, we obtain an estimate on the emission direction of the direct signals by considering the limit of large pulse durations $\omega\tau\gg 1$ and weak focusing $\omega w_0 \gg 1$. In this parameter regime, one can expect the signal photons of energy $k^0\approx 3\omega$ to be emitted in the vicinity of the direction $\hat{\vec{k}}_{\rm max}^{\rm pw} = {(m\hat{\vec{k}}_1 +(3-m) \hat{\vec{k}}_2)/|m\hat{\vec{k}}_1 +(3-m) \hat{\vec{k}}_2|}$ with $m\in\{1,2\}$.\\
Accordingly, in this plane wave limit, the polar and azimuthal angles of the emission direction maximizing the amplitude of $\frac{{\rm d}^2 N^{3\omega}}{{\rm d}\varphi {\rm d}\! \ct}$ are 
\begin{equation}
	\begin{aligned}
	\varphi_{\rm max}^{\rm pw}=&0 \, ,\\
	\vartheta_{\rm max}^{\rm pw}=&\arcsin\left(\frac{(2m-3)\stch}{\sqrt{5+4\ctc}}\right)
	\end{aligned}
	\label{eq:angles_plane_wave}\, .
\end{equation}
We emphasize that $\varphi_{\rm max}=0$ is expected to hold true also beyond the plane wave limit and for the interference terms ($p\neq q$ in Eq.~\eqref{eq:d2N_3omega}) since the field profiles \eqref{eq:E} introduce momentum components perpendicular to the collision plane only symmetrically. On the other hand, we expect deviations from $\vartheta_{\rm max}^{\rm pw}$ when taking the full spatial and temporal beam profile into account. 

\subsection{Analytical scalings}
\label{sec:3omega_scaling}
The difficulty of the integration over $\vartheta$ and $\varphi$ of Eq.~\eqref{eq:d2N_3omega} arises from the trigonometric dependencies on these angles, particularly in the exponent. As will be detailed below, this can be bypassed by approximating the signal with a function of Gaussian form, i.e. eliminating the dependence on the angles in the prefactor by substituting them with constants and expanding the argument of the exponent up to $\mathcal{O}(\vartheta^2)$ and $\mathcal{O}(\varphi^2)$, respectively.\\
First, we want to perform the integration over $\varphi$. To this end, we perform the expansion around $\varphi=0$ in both the overall prefactor and the arguments of the exponential functions. In the prefactor we keep only the leading term while in the exponent we keep contributions up to quadratic order in $\varphi$. This is motivated by the fact that the maximum of the differential signal photon number, Eq.~\eqref{eq:d2N_3omega}, can be found at this angle. As the partial signals are localized in a small angular region, it is possible to formally extend the integration limits to the complete real domain, leading to an integral of Gaussian form.
The resulting expression after the $\varphi$-integration, ${\rm d} N^{3\omega}/{\rm d}\!\st$, is rather lengthy without providing any additional insight and therefore not given explicitly.\\
We proceed similarly for the integration of ${\rm d} N^{3\omega}/{\rm d}\!\st$ over $\vartheta$: In a first step, we expand the argument of the exponential function around $\vartheta=0$ up to $\mathcal{O}(\vartheta^2)$. The validity of this approach is initially only apparent for the indirect contributions, which maximize for $\vartheta = 0$.
Notably, for the direct terms ($p=q$) the expansion of the exponential function around $\vartheta=0$ features a term linear in $\vartheta$, which causes a shift of the maximum of the exponential function to
\begin{equation}
	\begin{aligned}
		\vartheta_{\rm shift}^{p,q} = \frac{3\tau^2 \tilde{H}_1 \Theta_{pq} \stch}{\tilde{H}_1 \tilde{H}_2 -18 \tau^4 w_0^2 \sin^2(\frac{\vartheta_{\rm coll}}{2})\Theta_{pq}^2}
		\label{eq:3omega_th-shift}
	\end{aligned}
\end{equation}
with 
\begin{equation}
	\begin{aligned}
	\tilde{H}_1 =& w_0^2 \sin^2\left(\frac{\vartheta_{\rm coll}}{4}\right) (8H+\tau^2)+2\tau^2 H \cos^2\left(\frac{\vartheta_{\rm coll}}{4}\right) \, ,\\
	\tilde{H}_2 =& 20\tau^2 \cos^2\left(\frac{\vartheta_{\rm coll}}{4}\right) -(8H +\tau^2) \, .
	\end{aligned}
\end{equation}
For strongly focused beams with $\tau \gg w_0$, Eq.~\eqref{eq:3omega_th-shift} can be approximated by the simpler expression $\vartheta_{\rm shift} \approx 3 \stch /(5+10\ctch - 4 \ctc)$, whose range of values well below 1 justifies the expansion around $\vartheta=0$ also for the direct contributions. In addition, based on our estimations on the emission direction in the plane wave limit given in Eq.~\eqref{eq:angles_plane_wave} together with the expectation of a small optimal collision angle, the expansion in $\vartheta$ should be justified and give reasonable results also for weakly focused beams.\\
The maximum of the differential number of signal photons is expected to be close to $\vartheta=\vartheta_{\rm shift}^{p,q}$. Therefore, we set $\vartheta=\vartheta_{\rm shift}^{p,q}$ in the prefactor of the expression for ${\rm d} N^{3\omega}/{\rm d}\!\st$.
After integrating over $\vartheta$, we obtain the total number of $3\omega$-signal photons
\begin{equation}
	\begin{aligned}
		N^{3\omega} \approx & \left(\frac{7}{45}\right)^2 \frac{\alpha^4}{m_e^8} \frac{2^7 \sqrt{3}}{\pi^2} \frac{\omega W^3}{\tau^3 w_0^2} \left(\frac{\tilde{H}_1}{H}\right)^\frac{3}{2} \sin\left(\frac{\vartheta_{\rm coll}}{4}\right)\stch \\
		&\times \sum_{p,q=1,2} \frac{\prod_{m=p,q}\left[1-\cos\left(\vartheta_{\rm shift}^{p,q}-(-1)^m \frac{\vartheta_{\rm coll}}{2}\right) \right]}{\sqrt{\tilde{H}_1 \tilde{H}_2-18\tau^4 w_0^2 \sin^2 (\frac{\vartheta_{\rm coll}}{2}) \Theta_{pq}^2}}  \cos\vartheta_{\rm shift}^{p,q} \\
		&\times \frac{\sqrt{ 2H(4w_0^2 \mathfrak{h}^2 +\tau^2 \sin^2(\frac{\vartheta_{\rm coll}}{2})) + \tau^2 w_0^2 \left(\mathfrak{a}_+^2 + 9\mathfrak{a}_-^2 -6\Theta_{pq} \mathfrak{a}_+ \mathfrak{a}_- \right) }}{\sqrt{-8H\left(\mathfrak{a}_+^2 - \mathfrak{a}_-^2 + \mathfrak{a}_+ \right) -\tau^2 \left[\mathfrak{a}_+^2 + 9\mathfrak{a}_-^2 +\mathfrak{a}_+ -3\Theta_{pq} (2\mathfrak{a}_+ +1)\mathfrak{a}_+ \right]}} \\
		&\times \exp\Bigg\{-\frac{3\omega^2 \tau^2 w_0^2}{8\tilde{H}_1} \frac{\sin^2(\frac{\vartheta_{\rm coll}}{4})}{\tilde{H}_1 \tilde{H}_2 -18\tau^4 w_0^2 \Theta_{pq}^2 \sin^2(\frac{\vartheta_{\rm coll}}{2})} \bigg\{(\tau^2+8H)\tilde{H}_1 \tilde{H}_2 \\
		&\hspace{4cm}-36\tau^4 \Theta_{pq}^2 \cos^2(\frac{\vartheta_{\rm coll}}{4}) \left[2\tilde{H}_1 -3\tau^2 H \cos^2(\frac{\vartheta_{\rm coll}}{4})\right] \bigg\}\Bigg\} 
	\end{aligned}
	\label{eq:N_3omega_series-approx}
\end{equation}
with
\begin{equation}
	\begin{aligned}
		\mathfrak{a}_+ =& \frac{1}{2}\left.\left(h_+ + h_-\right)\right|_{\varphi=0, \vartheta=\vartheta_{\rm shift}^{p,q}} = \cos\vartheta_{\rm shift}^{p,q} \ctch -1 \\
		\mathfrak{a}_- =&\frac{1}{2}\left.\left(h_+ - h_-\right)\right|_{\varphi=0, \vartheta=\vartheta_{\rm shift}^{p,q}} = \sin\vartheta_{\rm shift}^{p,q} \stch\\
	\end{aligned}
\end{equation}
and
\begin{equation}
	\mathfrak{h} = h|_{\varphi=0, \vartheta=\vartheta_{\rm shift}^{p,q}} = \cos\vartheta_{\rm shift}^{p,q}- \ctch \, .
\end{equation}
To better display the dependence of the number of signal photons on the laser parameters, in Eq.~\eqref{eq:N_3omega_series-approx} we expressed the peak field amplitude $\mathfrak{E}$ in terms of the laser pulse energy via Eq.~\eqref{eq:PeakFieldAmplitude}.\\
Eq.~\eqref{eq:N_3omega_series-approx} simplifies considerably if we expand prefactor and exponent separately to leading order in $\vartheta_{\rm coll}$, resulting in 
 \begin{equation}
	N^{3\omega} \approx \left(\frac{7}{45}\right)^2 \frac{\alpha^4}{m_e^8} \frac{\omega W^3}{\tau^2 w_0^2} \frac{\vartheta_{\rm coll}^6}{11^3 \sqrt{33}} \sum_{p,q=1,2} \frac{(121-57\Theta_{pq}^2)^2 }{\sqrt{121-13\Theta_{pq}^2}}{\rm e}^{-\frac{27 \omega^2 w_0^2}{2816}(11-2\Theta_{pq}^2)\vartheta_{\rm coll}^2} \, .
	\label{eq:N_3omega_approx}
\end{equation}
This approach is motivated by the good agreement of the full expressions and the respective leading order expansions in the relevant parameter ranges of pulse durations $20 \text{ fs}\lesssim \tau \lesssim 150$ fs and of beam waists $0.02 \text{ }\mu\text{m} \lesssim w_0 \lesssim 2$ $\mu$m. The deviations are below $15\%$ for collision angles $\vartheta_{\rm coll} \lesssim 65^\circ$. Since the optimal collision angle is found precisely in this angle regime (see below), Eq.~\eqref{eq:N_3omega_approx} is expected to give accurate predictions on the maximal number of merged signal photons attainable in the collision of two laser pulses.\\
From Eq.~\eqref{eq:N_3omega_approx}, we can directly read of that $N^{3\omega} \sim \tau^{-2}$ and $N^{3\omega} \sim W^3$.
For the beam waist $w_0$ the scaling depends on the collision angle $\vartheta_{\rm coll}$ and the photon energy $\omega$. The larger $\vartheta_{\rm coll}$ and $\omega$, the fewer signal photons will be emitted when increasing $w_0$.\\
In turn, the scaling with $\omega$ depends on the collision angle $\vartheta_{\rm coll}$ and the beam waist $w_0$. On the one hand, the prefactor of Eq.~\eqref{eq:N_3omega_approx} increases when $\omega$ is increased. On the other hand, the exponential function causes the signal photon number to decrease with growing $\omega$. The last effect is larger, the larger $\vartheta_{\rm coll}$ and $w_0$, and for realistic laser configurations typically outweighs the enlarging effect of the prefactor. However, laser beams with larger photon energy can also be focused to smaller beam waists, such that the beam divergence $\theta = 2/(w_0 \omega)$ can be kept constant over a large range of $\omega$. For fixed $\theta$, it is beneficial for the signal to use large photon energies $\omega$.\\
From Eq.~\eqref{eq:N_3omega_approx} we also infer that the optimal collision angle $\vartheta_{\rm coll}^{\rm max}$ moves to smaller values when $w_0$ or $\omega$ are increased.
It even allows us to derive a formula for the optimal collision angle $\vartheta_{\rm coll}^{\rm max, pq}$ for each partial signal, which reads
\begin{equation}
	\vartheta_{\rm coll}^{\rm max, pq} = \frac{16\sqrt{11}}{3\omega w_0 \sqrt{11-2\Theta_{pq}^2}} \, .
	\label{eq:thc_3omega}
\end{equation} 
We obtained this result by finding the root of each summand in Eq.~\eqref{eq:N_3omega_approx} separately. Actually we are interested in the collision angle $\vartheta_{\rm coll}$, which maximizes the total number of signal photons. However, adopting the same procedure for the total number of signal photons, i.e. for all summands together, leads to an expression which is not analytically solvable for $\vartheta_{\rm coll}$. Nevertheless, Eq.~\eqref{eq:thc_3omega} is very useful to take an educated guess on the optimal collision geometry because $\vartheta_{\rm coll}^{\rm max, pq}$ for $p=q$ and $p \neq q$ do not differ very much. In fact, the product $\omega w_0$ has a lower bound due to the diffraction limit. Assuming $w_0 = \lambda$ this bound becomes $\omega w_0 = 2\pi$ and we can estimate that the optimal collision angles for the direct and the indirect signal are at most $\sim 5^\circ$ apart.
%The optimal collision angle $\vartheta_{\rm coll}^{\rm max}$ has to be in between the angles given by Eq.~\eqref{eq:thc_3omega} for $p=q$ and $p\neq q$.
For $w_0 = \lambda$, the optimal collision angle is at $\vartheta_{\rm coll}^{\rm max, p=q}\approx 53.8^\circ$ ($\vartheta_{\rm coll}^{\rm max, p\neq q} \approx 48.6^\circ$) for the direct (indirect) signal. For beams with larger beam waist $w_0$ the optimal collision angle is at lower values.\\
At the angles given by Eq.~\eqref{eq:thc_3omega} the exponential dependency of the signal photon number on the laser parameters drops out and we have 
		\begin{equation}
			\begin{aligned}
			N^{3\omega}_{{\rm max}} \approx& \left(\frac{7}{45}\right)^2 \frac{\alpha^4}{m_e^8} \frac{W^3}{\tau^2 \omega^5 w_0^8} \frac{2^{24}}{9^3 \sqrt{33}} {\rm e}^{-3} \left[1+\left(\frac{2^{11}\sqrt{3}}{9^4} -1\right)\Theta_{pq}^2\right] \\
			\approx & 4.83\frac{\alpha^4}{m_e^8} \frac{W^3}{\tau^2 \omega^5 w_0^8} \left[1-0.46\Theta_{pq}^2\right] \, .
			\end{aligned}
	\end{equation}
The scaling of the number of merged photons with the various laser parameters becomes particularly evident from this expression.

\section{Results}
\label{sec:results}
In the following we present example results of the $3\omega$-signal for laser parameters from from the high-power laser system (HPLS) available at ELI-NP \cite{ELI-NP, ELI_web} providing two identical laser pulses of photon energy $\omega=1.51$ eV, duration $\tau^{\rm HM}=24$ fs and energy $W=244$J at a repetition rate of $1/60$ Hz. Moreover, we assume these lasers to be focused to $w_0 = \lambda$.

\begin{figure}[htbp]
	\begin{minipage}{0.495\textwidth}
		\begin{subfigure}{\textwidth}
			\centering
			\includegraphics[width=\linewidth]{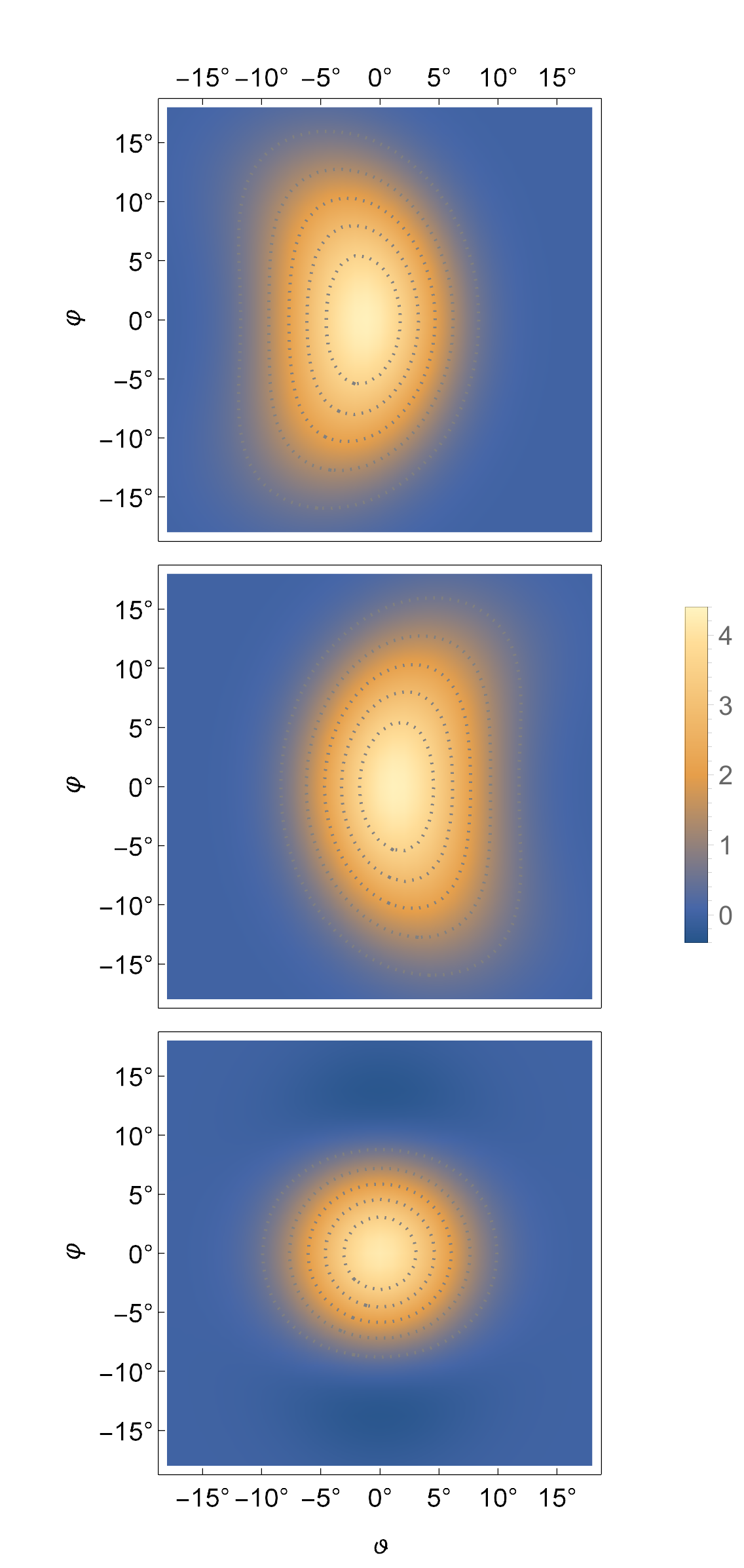} 
			\caption{Partial contributions}
			\label{fig:spatialDistribution_ELI-ELI_partial}
		\end{subfigure}
	\end{minipage}
	\begin{minipage}{0.495\textwidth}
		\centering
		\begin{subfigure}{0.97\textwidth}
			\centering
			\vspace{0.68cm}
			\includegraphics[width=\linewidth]{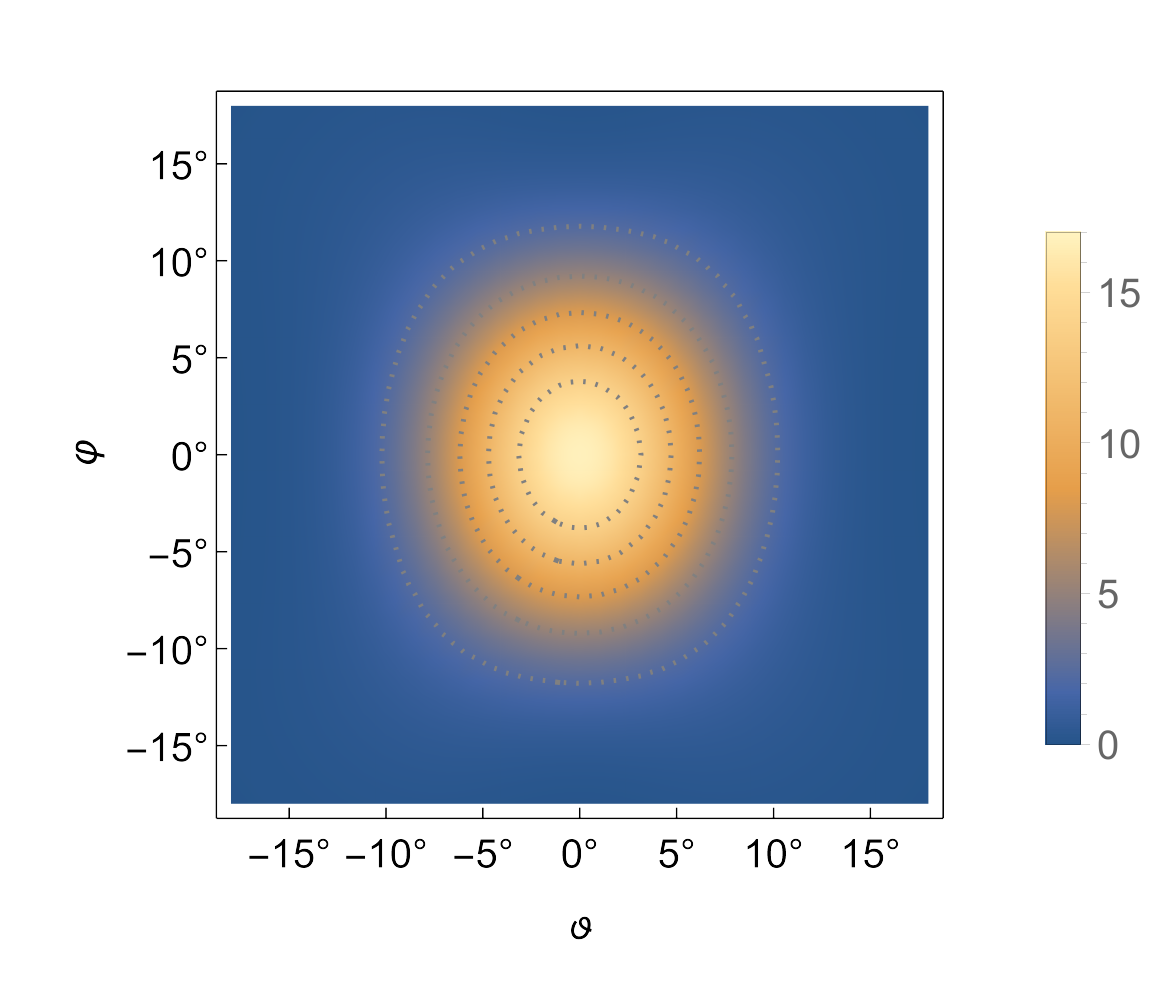}
			\caption{Total signal}
			\label{fig:spatialDistribution_ELI-ELI_total}
		\end{subfigure}
	\end{minipage}
	\caption{Angular distribution for the collision of two HPLS pulses according to Eq.~\eqref{eq:d2N_3omega} at $\vartheta_{\rm coll}=\vartheta_{\rm coll}^{\rm max, num} \approx 50.2^\circ$. Dashed lines indicate contours with the same differential contribution, with the outermost line being at $1/e^2$ of the maximum value ${\rm d}^2 N^{3\omega}_{\rm max}/({\rm d}\varphi {\rm d}\!\sin\vartheta)$. Left (from top to bottom): partial contributions for $p=q=1$, $p=q=2$ (both direct terms), and $p\neq q$ (interference term). Right: total signal (partial contribution of interference term included twice). Note the different scales for the partial contributions and the total signal.}
	\label{fig:spatialDistribution_ELI-ELI}
\end{figure}

We first discuss the spatial distribution of the signal photons in the collision of two identical optical laser beams at the numerically determined, optimal collision angle $\vartheta_{\rm coll}^{\rm max, num} \approx 50.2^\circ$.
Eq.~\eqref{eq:d2N_3omega} allows us to illustrate the contributions from the underlying microscopic processes separately, i.e. one photon of beam 1 merging with two photons of beam 2 (top panel of Fig.\ref{fig:spatialDistribution_ELI-ELI_partial}), two photons of beam 1 merging with one photon of beam 2 (middle panel of Fig.\ref{fig:spatialDistribution_ELI-ELI_partial}) and the interference of both (bottom panel of Fig.\ref{fig:spatialDistribution_ELI-ELI_partial}). As expected, the direct contributions (top, middle) are mirror images of each other with the mirror plane at $\vartheta=0$.
From Fig.~\ref{fig:spatialDistribution_ELI-ELI_partial} we can read off that the main emission directions of the direct contributions at this collision angle are only slightly shifted away from the angle bisector of the two incident beams to $\vartheta \approx \pm 1.5^\circ$, respectively. The angular extent of the signal components is much larger: The direct contributions drop to $1/e^2$ of their maximal amplitude at $\Delta \vartheta \approx 10.5^\circ$ and $\Delta \varphi \approx 15.5^\circ$ away from their main emission directions. Consequently, the contribution of the interference term is almost as large as the direct contribution at its main emission direction, i.e. at $\vartheta = 0 = \varphi$. One has to keep in mind, that the interference term as shown in the bottom panel of Fig.~\ref{fig:spatialDistribution_ELI-ELI_partial} contributes twice to the total signal. Thus, the interference is responsible for a fraction of about one third to one half of the total number of signal photons at the optimal collision angle. Note however, that the interference term can also reduce the number of signal photons in some angular regions. In the case shown here, the contribution of the interference term is slightly negative around $\vartheta \approx 0^\circ$ and $\varphi \approx \pm 12^\circ$.\\
Although it is very interesting to understand how the single terms of Eq.~\eqref{eq:d2N_3omega} contribute, they can of course never be isolated in an experiment because the signal photons associated with the different contribution are indistinguishable, and only the total amplitude squared is physical and strictly positive. Therefore, one will always only measure the total signal shown in Fig.~\ref{fig:spatialDistribution_ELI-ELI_total}. The main emission direction of the total signal is found at $\vartheta= 0 = \varphi$ as long as the direct contributions are not too far apart. Otherwise, the total signal will exhibit two main emission directions at $\varphi = 0$ and some $\vartheta = \pm \vartheta_{\rm max}$. In the particular case considered here, the total signal is almost circularly distributed with $1/e^2$-widths of about $\Delta \vartheta \approx 10^\circ$ and $ \Delta \varphi \approx 12^\circ$. For comparison note that the full beam divergences of the driving laser beams are given by $\Theta=2/\pi\approx 36^\circ$.

\begin{figure}
	\centering
	\includegraphics[width=0.7\linewidth]{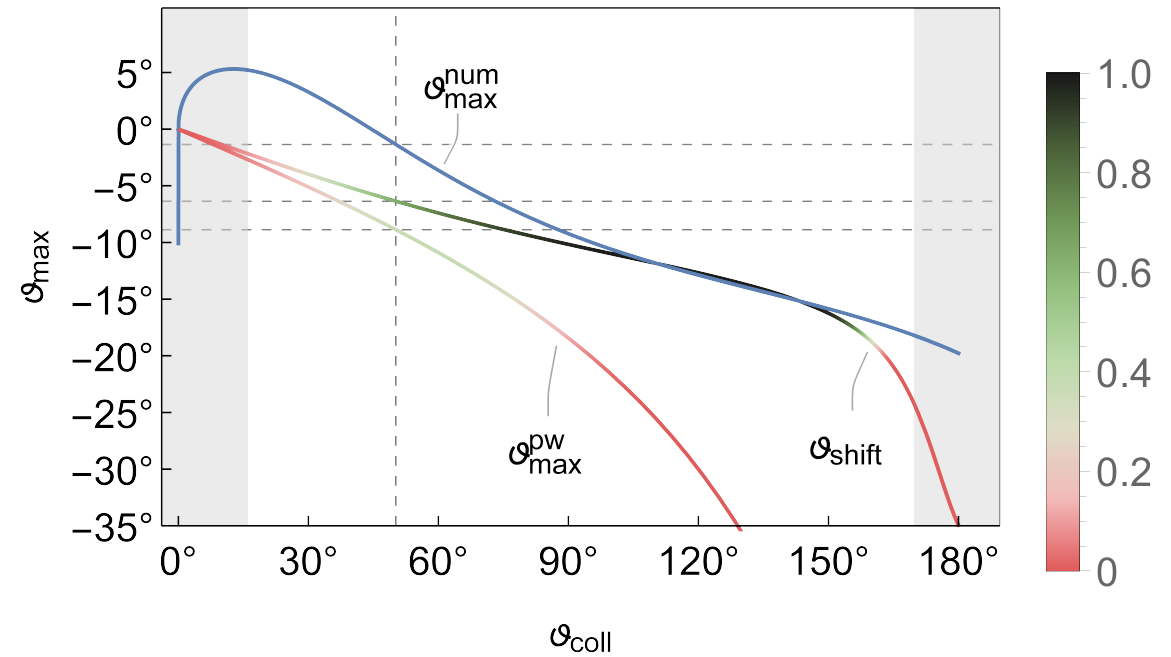}
	\caption{Comparison of solutions $\vartheta_{\rm max}^{\rm pw}$ (Eq.~\eqref{eq:angles_plane_wave}) and $\vartheta_{\rm shift}^{p,q}$ (Eq.~\eqref{eq:3omega_th-shift}) for the direction of maximal emission of the partial contribution with $p=1=q$ with the numerical solution $\vartheta_{\rm max}^{\rm num}$ obtained by numerically maximizing the $(p=1=q)$-contribution of Eq.~\eqref{eq:d2N_3omega} for a collision angle of $\vartheta_{\rm coll}$. The curves of the approximate solutions are colored according to $\left(\frac{{\rm d}^2 N^{3\omega}}{{\rm d}\varphi {\rm d}\sin\vartheta}\right)\big|_{\vartheta=\vartheta_{\rm max}^{\rm approx}}/\left(\frac{{\rm d}^2 N^{3\omega}}{{\rm d}\varphi {\rm d}\sin\vartheta}\right)\big|_{\vartheta=\vartheta_{\rm max}^{\rm num}}$ with $\vartheta_{\rm max}^{\rm approx}=\vartheta_{\rm max}^{\rm pw}$ and $\vartheta_{\rm max}^{\rm approx}=\vartheta_{\rm shift}^{1,1}$, respectively. The vertical dashed line marks the numerically determined optimal collision angle $\vartheta_{\rm coll}^{\rm max, num}$. The horizontal dashed lines mark the corresponding directions of maximum emission predicted by the different approaches, i.e.  $\vartheta_{\rm max}^{\rm pw}$, $\vartheta_{\rm shift}^{1,1}$ and $\vartheta_{\rm max}^{\rm num}$. The gray areas mark where the infinite Rayleigh range approximation breaks down.}
	\label{fig:emissionDirection_opt-opt_thc}
\end{figure}

While the direction of maximal emission of the indirect contribution is known to be $\vartheta_{\rm max} = 0 =\varphi_{\rm max}$ for symmetry reasons, the emission directions of the direct contributions depend on the collision angle and the laser parameters. Therefore, we now analyze the emission directions of the direct contributions in more detail with special focus on their dependence of the chosen collision angle. To this end, we numerically determine the direction of maximal emission $\vartheta_{\rm max}^{\rm num}$ for the partial contributions separately from Eq.~\eqref{eq:d2N_3omega}. With this as a reference, we test the quality of our approximation of the emission direction $\vartheta_{\rm max}^{\rm pw}$ deduced from plane wave considerations, Eq.~\eqref{eq:angles_plane_wave}, and also that of $\vartheta_{\rm shift}^{p,q}$ as given in Eq.~\eqref{eq:3omega_th-shift}.
Without loss of generality we consider only the partial contribution with $p=1=q$, which gives us Fig.~\ref{fig:emissionDirection_opt-opt_thc}. The second direct contribution ($p=2=q$) would give us the same picture but mirrored at $\vartheta_{\rm max} = 0$. \\
The direction of maximal emission $\vartheta_{\rm max}^{\rm num}$ exhibits several interesting features. Firstly, it is much closer to the angle bisector between the two incident beams ($\vartheta=0$) than one might have expected from the plane wave considerations, i.e. compared to $\vartheta_{\rm max}^{\rm pw}$. Secondly, for small collision angles, the signal photons are even scattered closer towards the beam contributing only with one photon than to the beam contributing with two photons. At a collision angle of $\vartheta_{\rm coll}\approx 45.5^\circ$, this behavior is reversed. The appearance of this extremum in the curve of the direction of maximal emission hints to the existence of two opposing effects.
The second suggested expression for the direction of maximal emission, $\vartheta_{\rm shift}^{p,q}$, approximates the numerical determined curve better than the plane wave solution $\vartheta_{\rm max}^{\rm pw}$ but also fails to reproduce this essential feature. For a weaker focusing of the laser beams $\vartheta_{\rm max}^{\rm num}$ as well as $\vartheta_{\rm shift}^{p,q}$ converge towards $\vartheta_{\rm max}^{\rm pw}$.\\ 
To not only give an impression on the deviation of the predicted angles of maximal emission but also on the corresponding signal amplitudes, we encode the deviation in the differential signal photon numbers at the angles $\vartheta_{\rm max}^{\rm pw}$ and $\vartheta_{\rm shift}^{p,q}$ compared to that at $\vartheta_{\rm max}^{\rm num}$ in the color of the curves, i.e. the color is given by $\left(\frac{{\rm d}^2 N^{3\omega}}{{\rm d}\varphi {\rm d}\sin\vartheta}\right)\big|_{\vartheta=\vartheta_{\rm max}^{\rm approx}}/\left(\frac{{\rm d}^2 N^{3\omega}}{{\rm d}\varphi {\rm d}\sin\vartheta}\right)\big|_{\vartheta=\vartheta_{\rm max}^{\rm num}}$ with $\vartheta_{\rm max}^{\rm approx}=\vartheta_{\rm max}^{\rm pw}$ and $\vartheta_{\rm max}^{\rm approx}=\vartheta_{\rm shift}^{p,q}$, respectively.\\
As we are interested in getting the signal as large as possible, we pay special attention to the properties at the optimal collision angle $\vartheta_{\rm coll}^{\rm max, num} \approx 50.2^\circ$, marked with dashed lines in Fig.~\ref{fig:emissionDirection_opt-opt_thc}. Here, the different approaches predict $\vartheta_{\rm max}^{\rm num} = -1.4^\circ$, $\vartheta_{\rm shift}^{1,1} = -6.4^\circ$ and $\vartheta_{\rm max}^{\rm pw} = -8.9^\circ$. Evaluating the differential signal photon number at $\vartheta_{\rm shift}^{1,1}$ and $\vartheta_{\rm max}^{\rm pw}$ instead of  $\vartheta_{\rm max}^{\rm num}$ leads to a $37\%$, respectively $65\%$, smaller signal.
Despite these discrepancies, $\vartheta_{\rm shift}^{p,q}$ and $\vartheta_{\rm max}^{\rm pw}$ serve as valid points around which to integrate numerically in order to obtain the total number of signal photons. Furthermore, we can use these approximations for theoretical purposes, e.g. to be able to determine an approximate expression for the total number of signal photons analytically, as done in Sec.~\ref{sec:3omega_scaling}. 
\\

\begin{figure}
		\begin{minipage}{0.49\textwidth}
			\begin{subfigure}{\textwidth}
				\centering
				\includegraphics[width=\linewidth]{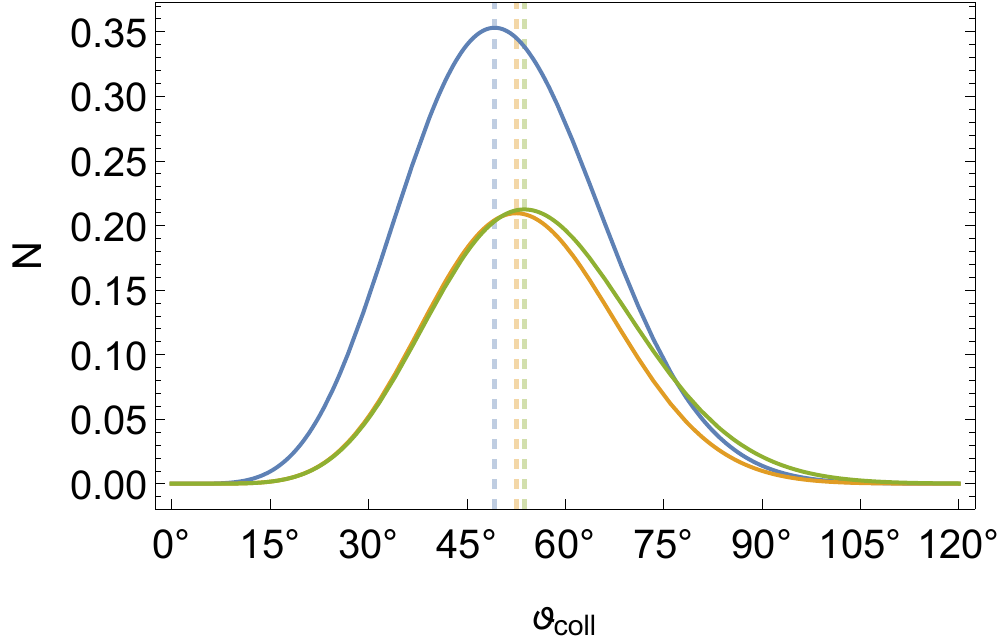}
				\caption{Direct contribution}
				\label{fig:signalphotonnumber_ELI_direct}
			\end{subfigure}\\ \vspace{15pt}
			\begin{subfigure}{\textwidth}
				\centering
				\includegraphics[width=\linewidth]{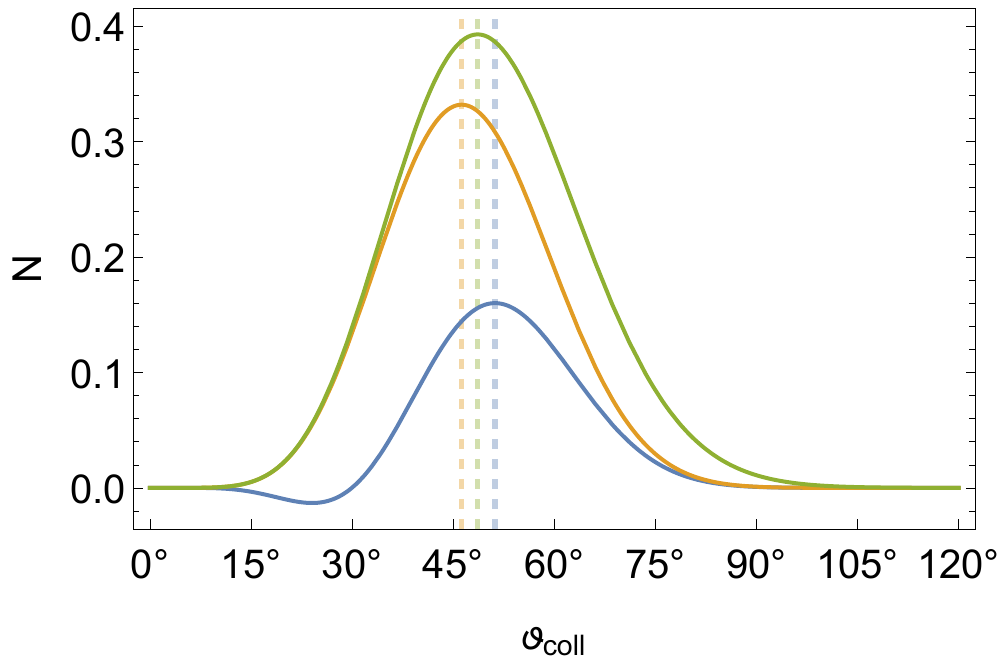} 
				\caption{Indirect contribution}
				\label{fig:signalphotonnumber_ELI_indirect}
			\end{subfigure}
		\end{minipage}
		\hfill
		\begin{minipage}{0.49\textwidth}
			\begin{subfigure}{\textwidth}
				\centering
				\includegraphics[width=\linewidth]{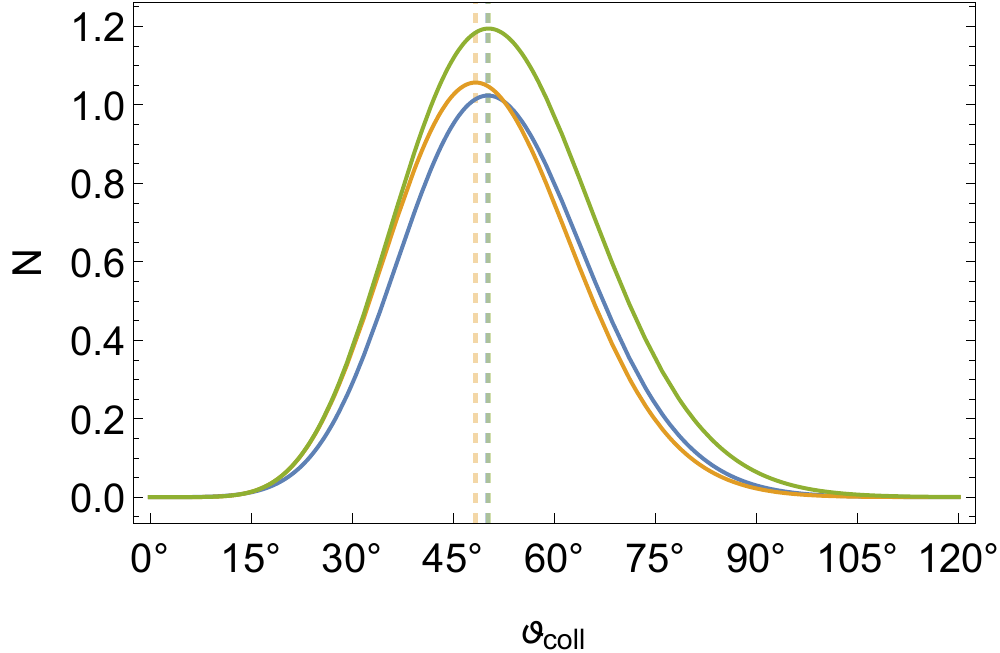}
				\caption{Total signal}
				\label{fig:signalphotonnumber_ELI_total}
			\end{subfigure}
		\end{minipage}
		\caption{Number of merged photons $N^{3\omega}$ as a function of the collision angle $\vartheta_{\rm coll}$ in a collision of two HPLS pulses for the direct contribution with $p=1=q$ (\subref{fig:signalphotonnumber_ELI_direct}), the indirect contribution with $p=1$, $q=2$ (\subref{fig:signalphotonnumber_ELI_indirect}) and the total signal (\subref{fig:signalphotonnumber_ELI_total}). The blue curves follow upon numerically integrating Eq.~\eqref{eq:d2N_3omega} over the emission angles $\vartheta$ and $\varphi$. The orange and green curves follow from the approximate expressions of Eqs.~\eqref{eq:N_3omega_series-approx} and \eqref{eq:N_3omega_approx}, respectively. The dashed vertical lines mark the corresponding predictions for the optimal collision angle $\vartheta_{\rm coll}^{\rm max}$, where the signal photon number is maximized.}
		\label{fig:signalphotonnumber_ELI}
	\end{figure}

We obtain the number of signal photons $N^{3\omega}$ as a function of the collision angle $\vartheta_{\rm coll}$ by numerically integrating Eq.~\eqref{eq:d2N_3omega}. In Fig.~\ref{fig:signalphotonnumber_ELI}, the numerically determined direct ($p=1=q$, Fig.~\ref{fig:signalphotonnumber_ELI_direct}), indirect ($p=1$, $q=2$, Fig.~\ref{fig:signalphotonnumber_ELI_indirect}), and total $3\omega$-signal (Fig.~\ref{fig:signalphotonnumber_ELI_total}) are presented together with the corresponding approximate results according to Eq.~\eqref{eq:N_3omega_series-approx} and Eq.~\eqref{eq:N_3omega_approx}.
	\\
While the direct contribution is being underestimated by the approximations in Eq.~\eqref{eq:N_3omega_series-approx} and Eq.~\eqref{eq:N_3omega_approx}, the indirect contribution is being overestimated. The discrepancy between the numerical result and the approximations for the indirect contributions are mainly related to the neglect of higher order term in the expansion in $\varphi$ and $\vartheta$ of ${\rm d}^2 N^{3\omega}/{\rm d}\varphi {\rm d} \!\st$. In the case of the direct contributions, there is also the fact that $\vartheta = 0$ is not optimal as an expansion point. As pointed out above, for a beam which is more plane-wave-like, the direction of emission can be well approximated by $\vartheta_{\rm max}^{\rm pw}$ which fulfills $\left|\vartheta_{\rm max}^{\rm pw}\right| > \left|\vartheta_{\rm max}^{\rm num}\right|$ for the parameters considered here. Therefore, the level of agreement between the numerical and the approximated results involving an expansion in $\vartheta\ll 1$ and thus also requiring $\vartheta_{\rm max} \ll 1$ are expected to worsen with increasing beam waist and pulse durations relatively to those presented in Fig.~\ref{fig:signalphotonnumber_ELI_direct}. Interestingly, the numerical and the approximate analytical results are in good agreement again for the total signal. Also, the position of the optimal collision angle is reproduced with a deviation of less than $2^\circ$ for the total signal. The deviations of the predicted optimal collision angle are larger for the partial contributions, but still below $5^\circ$. While the optimal collision angles predicted by the approximate expressions are shifted to larger values for the direct contribution, they are shifted to lower values for the indirect contribution.\\
In summary, for the specific laser parameters stated at the beginning of this section we find the maximal number of merged photons to be 1.02 per shot, or, taking the repetition rate of $1/60$ Hz into account, 61.2 per hour. The indirect contribution contributes approximately one third of the total signal photons. The collision angle under which these numbers can be reached is $\vartheta_{\rm coll} \approx 50.2^\circ$. 

\section{Conclusions and Outlook}
\label{sec:concls}
In the present work we have studied laser photon merging in the collision of two identical laser beams, modeled as pulsed Gaussian beams in the infinite Rayleigh range approximation. We have derived approximate expressions for the optimal collision angle and the signal photon number of the $3\omega$-signal and have compared them to numerical results. Within their limits of validity our approximations give consistent results compared to numerical calculations over a large parameter range. For the example of the collision of two optical high-intensity lasers of the 10 PW class available at ELI-NP, at zero impact the total signal photon number was reproduced to an accuracy of $\approx 20\%$. This suggests hat our analytical approximations provide a convenient way to analyze the $3\omega$-signal with little numerical effort.\\
We have found the attainable number of merged photons per hour at the optimal collision angle of $\vartheta_{\rm coll}\approx50.2^\circ$ to be $\approx 61.2$ at a repetition rate of $1/60$ Hz for these parameters.
Note also that for 1 PW class laser facilities the merging signal is still sizable. E.g. for the parameters of the Center of Advanced Laser Applications (CALA) in Munich ($\omega = 1.55$ eV, $\tau^{\rm HM}=20$ fs ,$W=24$ J and $w_0 =800$ nm) \cite{CALA_web, Hartmann2021}, one can expect to obtain about $5.5$ merged photons per hour. The considerably lower laser energy is here compensated by the higher repetition rate of the laser of $1$ Hz. Although these numbers are much lower than the signal photon numbers associated with vacuum birefringence, the merging signal can still be considered as prospective for a discovery experiment of quantum vacuum nonlinearity. Its clear advantage is that the frequency $3\omega$ signal photons can be clearly distinguished from the driving laser beams.
In contrast, the signal of vacuum birefringence typically competes with the large background of laser photons of the same frequency as the signal. Correspondingly, the discernible number of signal photons can be considered to be of the same order as the merging signal for the respective best options of collision geometry and laser systems available today, cf. e.g. \cite{Mosman2021}.
Note also that three beam scenarios as discussed e.g. recently in Ref.~\cite{Gies:2017ezf, King:2018wtn, Gies2021} provide much larger merging signals than the two beam scenario discussed here. However, the experimental implementation of a collision of three beams at a well defined spacetime point is considerably more difficult than the already extremely demanding task of colliding two tightly focused high-intensity laser beams.\\
In the present study we have emphasized the perspective of the nonlinear quantum vacuum signature of laser photon merging in a two-beam configuration. We have identified interesting features of the merging signal such as the acute optimal collision angle and the considerable number of signal photons. With the quantum vacuum signal of two-beam laser photon merging we have revealed a further possibility for the discovery of QED vacuum nonlinearities in an all-optical experiment. Most notably our proposal requires the collision of only two fundamental-frequency laser beams and thus avoids many experimental complications inherent to collision scenarios involving three or more high-intensity laser beams.

\acknowledgments

This work has been funded by the Deutsche Forschungsgemeinschaft (DFG) under Grant No. 416607684 within the Research Unit FOR2783/2.

\end{document}